**Title**: Identifying Influential Spreaders of Epidemics on Community Networks
**Authors**: Shi-Long Luo, Kai Gong* and Li Kang
**Affiliations**: School of Economic Information Engineering, Southwestern University of Finance and Economics, Chengdu, Peoples Republic of China
**Correspondence author**:
Kai Gong, PhD
Southwestern University of Finance and Economics
555, Liutai Avenue, Wenjiang District, Chengdu, Sichuan, P. R. China, 611130.
E-mail: gongkai1210@swufe.edu.cn

## ABSTRACT

An efficient strategy for the identification of influential spreaders that could be used to control epidemics within populations would be of considerable importance. Generally, populations are characterized by its community structures and by the heterogeneous distributions of weak ties among nodes bridging over communities. A strategy for community networks capable of identifying influential spreaders that accelerate the spread of disease is here proposed. In this strategy, influential spreaders serve as target nodes. This is based on the idea that, in k-shell decomposition, weak ties and strong ties are processed separately. The strategy was used on empirical networks constructed from online social networks, and results indicated that this strategy is more accurate than other strategies. Its effectiveness stems from the patterns of connectivity among neighbors, and it successfully identified the important nodes. In addition, the performance of the strategy remained robust even when there were errors in the structure of the network.

## 1. INTRODUCTION

Epidemics can lead to serious loss of life and they have huge an impact on the economy (Keeling and Rohani 2008), as witnessed during the 2003 outbreak of severe acute respiratory syndrome (SARS) (Gomez-Gardenes et al. 2008; Colizza et al. 2006), the 2009 outbreak of H1N1 influenza A virus (Fraser et al. 2009), and the 2013 outbreak of H7N9 Influenza A virus (Salathe et al. 2013). Knowledge regarding the pathways by which diseases spreading through networks and how this network might be used to prevent epidemics is of great importance. This issue has attracted a great deal of attention from researchers across various fields (Kitsak et al. 2010; Salathe and Jones 2010). Identifying the influential spreaders that can hinder the spread of disease effectively so as to suppress outbreaks remains an open issue (Ghoshal and Barabasi 2011).

Hubs, individuals who have high centrality in networks, are commonly believed to be the most influential nodes in the spreading process because they can affect many neighbors (Pastor-Satorras and Vespignani 2001; Cohen et al. 2001; Goh et al. 2003). In the case of networks with broad-degree distribution (Barabasi and Albert 1999), the *degree* strategy for the well-connected individuals has been shown to be an efficient method of identifying efficient spreaders (Cohen et al. 2001; Pastor-Satorras and Vespignani 2001). *Betweenness* is another centrality strategy. It involves measuring the number of shortest paths that cross the current node. It has been used to determine who has the most influence on others in networks (Freeman 1978; Goh et al. 2003). However, Kitsak et al. pointed out that the most efficient spreaders are those located within the core of a network as targeted by the *k-shell* decomposition strategy rather than targeted by degree and betweenness (Kitsak et al. 2010). This method is based on iterative pruning of nodes with degree smaller than or equal to the *k-core* index of the current layer until each node is associated with k-core index that reflects the core or periphery layer in network (Carmi et al. 2007).

Community structure (Fortunato 2010) is ubiquitous in complex networks (Girvan and Newman 2002), such as Facebook (Traud et al. 2011) and Twitter (Goncalves et al. 2011). It serves an important function in the dynamics of epidemic (Liu and Hu 2005; Wu and Liu 2008; Huang et al. 2006). In the presence of community structures, heterogeneous distribution of the number of weak ties was observed among real networks, such as air traffic networks (Guimera et al. 2005), social networks (Arenas et al.

2010), and communication networks (Onnela et al. 2007). The weak ties(Granovetter 1973) connecting a pair of nodes belonging to different communities have been found to provide shortcuts from one community to another (Gong et al. 2011). These ties have been proven to be more efficient in diffusing diseases through network (Hebert-Dufresne et al. 2013). Identifying the influential spreaders in community networks is quite challenging because the mesoscopic features of the community structure are complex (Gong et al. 2013). Here, a *k-shell with community* strategy is proposed. This strategy, based on the idea of the k-shell decomposition process, involves identifying the influential spreaders using weak ties and strong ties. Results demonstrated that the proposed strategy performs better in empirical networks than degree, betweenness, and k-shell decomposition strategies do. Simulation also shows that this strategy has the merit of being significantly robust against noise.

## 2. DATAS AND MODEL

The present paper compares results based on the current strategy and identifying efficient spreaders in empirical networks within epidemiological models. First, details are given with respect to the following issues: network construction and dynamic model.

### 2.1 Network Construction

Empirical networks are constructed using online collegiate social data from Facebook (https://code.google.com/p/socialnetworksimulation). Data from universities in U.S. was studied here. It includes anonymous data from students from Caltech, Princeton, Georgetown, and the University of Oklahoma. The data concern the dormitories, majors, and year for each individual, and dormitories were found to be key elements in the social organization of large universities (Traud et al. 2011). Based on these data, the networks were constructed by linking up pairs of individuals who (i) were online friends and lived in the same dormitory, or (ii) they lived in different dormitories but had the same major and year.

The giant network was then used for the present study. Basic statistical properties of networks are given in Table 1. All networks exhibited small-world characteristics with high clustering coefficient and short average path length. They also had high modularity.

## 2.2 Dynamic Model

In the real world, individuals will be able to contact only a limited number of people at once despite their wide acquaintance, and considerable examples have been observed: epidemic contact networks (Zhou et al. 2006), peer-to-peer distributed systems (Jovanovic 2001), and network marketing (Kim et al. 2006) . Here, the susceptible infected recovered (SIR) with identical capability of active contacts model was used to compare the performance of different identification strategies (Zhou et al. 2006). In the SIR model, each node in each network represented an individual who could be in one of three states: susceptible, infected, or recovered, and each link between nodes represented one connection that could spread an infection. Initially, all nodes were susceptible. To initiate an infection, one node was randomly chosen and considered infected. Each step and every individual had same infectivity $A = 2$, in which every infected individual generated identical contacts, multiple contacts to one neighbor were allowed, and the probability that a given susceptible node would be infected was $\beta = 0.5$. The probability that an infected node would recover was $\gamma = 1$. Once an individual was recovered, there would be no further change. In the simulation, states of every node were updated synchronously. The dynamics ended when all infected recovered. The average size of the infected, $M$, and the fraction of the population ever infected at the end of the epidemic were recorded, allowing quantification of the influence of given node on spreading process.

# 3. METHODS
## 3.1 Identification Strategies

The ideas behind degree, betweenness, and k-shell decomposition strategy are outlined, and the current strategy is discussed. Briefly, degree strategy was based on the idea that most influence nodes would be those with the largest number of connections, and it is one measure of local influence: only the structure around the node has to be considered (Latora and Marchiori 2007). Betweenness measures the number of shortest paths from all nodes to others that cross through that node. Kitsak et al. argues that the structure of network organization serves an important function such that there are plausible circumstances under which the most degree or highest betweenness as influential spreaders have the least pronounced impact on the spreading process (Kitsak et al. 2010). K-shell decomposition is one strategy based on iteratively pruning of nodes with degree no more than k-core index of the current layer. The highest k-core index is closely related to the concept of most influential nodes on spreading process. They used the strategy by identifying the core and periphery of given node in real network to identify key spreaders and found that k-shell decomposition strategy is more accurate. In fact, under real-world conditions, such as those in worldwide air traffic network and empirical networks (Gong et al. 2013; Guimera et al. 2005), the heterogeneous distribution in the number of weak ties among nodes indicated a pronounced difference in the spreading process. Unfortunately, this issue was not taken into account in the process of k-shell decomposition.

## 3.2 K-Shell with Community Strategy

The k-shell with community strategy is here proposed as a means of more effective identification. It is based on the idea that k-shell decomposition involves both weak ties and strong ties. Here, strong ties are those for which the source and target of the connection lie inside the same community. Figure 1 shows an example of weak ties and strong ties in simple community networks through visualization.

The k-shell with community strategy starts with successive pruning of the network by removing nodes with weak ties and strong ties separately. This process has three main components: (i) removal of nodes with weak ties, (ii) removal of nodes with strong ties, and (iii) assignment of k-score.

(i) After initialization of networks by removal of all nodes with strong ties, all nodes with weak ties $k_w = 1$ among nodes bridging over communities were then removed, and some nodes with weak ties may have remained, so we continued pruning the network repeatedly until there was no node left with $k_w = 1$ in the network. These nodes are associated with an index $k_{core}^W = 1$. In a similar step in the original work, the next level, $k_w = 2$, was iteratively removed and the system continued removing higher $k_w$ until all nodes were associated with an index of $k_{core}^W$.

(ii) After initialization of networks by removal of all weak ties, a procedure analogous to previous component was repeated by removing nodes with strong ties $k_s = 1$ until there were no nodes left with $k_s = 1$. These nodes are associated with an index of $k_{core}^S = 1$. The procedure was repeated until each node was associated with an index $k_{core}^S$.

(iii) Finally, the *k-score* was assigned to each node. It can be defined using the following equation:

$$\text{k-score} = (\alpha \times k_{core}^s) + \left[(1-\alpha) \times (1 + k_{core}^w)\right] \qquad (1)$$

The parameter of α giving one coefficient between 0 and 1, where 1 was preferred to hub nodes in local structure and 0 had a greater focus on bridge hubs, was shown in a previous work (K. Gong et al. 2013). The strategy is illustrated schematically in Figure 2.

## 4. RESULTS

### 4.1 Heterogeneity and Roles in Empirical Networks

To illustrate the necessity of an identification strategy that focuses on the most efficient spreaders in networks using heterogeneous distributions, the distributions of the cumulative probability density were analyzed are by weak ties and strong ties in real networks are denoted (Figure 3). The phenomenon of heterogeneity indicates that nodes have significantly different according to their pattern. This phenomenon also indicates the striking influence that heterogeneity has on the spreading process. As shown in Figure 4, roles of nodes by shapes in *Z-P* parameter space in empirical networks indicated that functional differences would emerge because of heterogeneity (Guimera and Amaral 2005). Here, *Z*, the intra-community score measured how many connections nodes have in the same community, and the participation coefficient *P*, often measures how well distributed the weak ties of nodes are among communities. These approach 1 if connections are uniformly distributed among all communities and 0 if are all within its vicinity. It can be defined using the following two equations:

$$Z = \frac{(k_s - <k_s>_g)}{\delta_g} \tag{2}$$

$$P = 1 - \sum (\frac{k_w^g}{k})^2 \tag{3}$$

Here, $k_s$ is the number of strong ties, $<k_s>_g$ is the average strong ties inside community $g$, $\delta_g$ is the standard deviation of $<k_s>_g$, and $k_w^g$ is the weak ties inside community $g$. $k$ is the total degree of node.

Figure 4 shows the following: (i) local hub ($Z \geq 2.5$) indicated that nearly all connections were local ($P < 0.2$) or across a few communities ($P < 0.6$); and (ii) sizeable normal nodes ($Z < 2.5$) are well distributed in the entire network ($0.6 \leq P < 0.8$). That will make it necessary for identify efficient spreaders with the heterogeneity.

## 4.2 Comparison of Spreading Efficiency

To compare the efficiency of degree, betweenness, and k-shell with that of our strategy, the SIR was performed with identical activity dynamics on empirical networks. The *imprecision function* $\varepsilon$, in a previous work(Kitsak et al. 2010), quantifies the difference between the average infected between the *fN* nodes (0 < $f$ < 1) with the highest degree, betweenness, k-core index, k-score, and the average infected of the $fN$ most efficient spreaders. For a given fraction $f$ the set of the $fN$ most efficient spreaders was first identified as measured by $M_{eff}$ (designated $F_{eff}$). Similarly, the $fN$ nodes with the highest k-score ($F_{k\text{-score}}$) were identified. In this way, the imprecision of k-score can be defined as follows:

$$\varepsilon_{k\text{-score}} = 1 - \frac{M_{k\text{-score}}}{M_{eff}} \qquad (4)$$

Here, $M_{k\text{-score}}$ and $M_{eff}$ are the average infected percentages over the $F_{k\text{-score}}$ and $F_{eff}$ sets of nodes, respectively. $\varepsilon_{k\text{-score}}$ approaches 0, an indication that the highest nodes are chosen by the strategy and are usually those that contribute the most to epidemics, and vice versa. $\varepsilon_{degree}$, $\varepsilon_{betweenness}$, and $\varepsilon_{k\text{-core}}$ are defined similarly to $\varepsilon_{k\text{-score}}$. Here, the differences $\Delta\varepsilon_i$ would indicate the spreading efficiency of strategy relative to the proposed strategy in the $fN$ set can be defined using the following equation:

$$\Delta\varepsilon_i = \frac{\varepsilon_i - \varepsilon_{k\text{-score}}}{\varepsilon_i} \qquad (5)$$

In most cases, the differences of $\Delta\varepsilon$ are positive over almost all of the set of different strategies (Figure 5). For example, k-score was on average 7.66% (40.43%, 55.06%) higher than k-core (degree, betweenness) for the Caltech data set, 5.28% (12.91%, 42.19%) higher for the Princeton set, 5.66% (17.88%, 41.65%) higher for the Georgetown set, and 4.93% (15.48%, 40.34%) higher for the University of Oklahoma set.

### 4.3 Comparison of Average Number of Connections among Neighbors

As spreaders, whether it has influence on spreading process is closely related to its pattern of connections in speeding up the transmission of epidemics. The effectiveness of k-score is illustrated in Figure 6, which compares the weak ties and strong ties among neighboring nodes identified using different strategies in networks. Here, the set $\Gamma$ is the union of neighbors of nodes those identified by strategy. Let $<k_w>(\Gamma_i)$, with $i$ denoting strategy, be the average number of weak ties within the union $\Gamma$ after applying said strategy. Similarly, $<k_s>(\Gamma_i)$ may be the average number of strong ties. The difference $\Delta<k_w>(\Gamma_i)$ and $\Delta<k_s>(\Gamma_i)$ are the measure of how effective k-score identify local hubs and bridge hubs within networks. This can be defined using the following two equations:

$$\Delta<k_w>\Gamma_i = <k_w>(\Gamma)_{k\text{-score}} - <k_w>(\Gamma)_i \quad (6)$$

$$\Delta<k_s>\Gamma_i = <k_s>(\Gamma)_{k\text{-score}} - <k_s>(\Gamma)_i \quad (7)$$

As shown in Figure 6, $\Delta > 0$ in almost all the cases indicates that nodes identified by k-score have more extensive connections of neighbors with well-connected than other strategies do, improving the effectiveness of identification for efficient spreaders.

## 4.4 Robustness of the Strategy

Another important aspect of an effective strategy is the robustness to missing or noise information. Such errors are common in real networks due to, for example, the inconsistency with which two individuals describe their relationship (Lu et al. 2011). Using empirical networks, the incomplete information by randomly removing the percentage of connections $f$, then have tested the robustness by using Kendall rank correlation coefficient $\tau(-1 \leq \tau \leq 1)$ (Kendall 1938), which measures the similarity of the ordering of nodes when ranked by k-score on the incomplete and original network. This can be defined using the following equation:

$$\tau = \frac{\text{number of concordant pairs} - \text{number of discordant pairs}}{N(N-1)/2} \quad (8)$$

$\tau$ is very similar to 1 denotes an exact agreement between two ranks for both elements, and vise-verse. Figure 7 shows that the proposed strategy performed stably at $\tau(>0.8)$ even $f$ larger than 20%.

## 5. CONCLUSIONS AND DISCUSSION

The heterogeneity distribution of weak ties under real-world conditions is important to identifying those nodes which are most influential spreaders in the transmission of diseases, but these nodes are difficult to identify in community networks. In summary, the k-shell with community strategy proposed here was studied for effectiveness. There is one strategy for identifying nodes. This strategy is based on the idea that k-shell decomposition involves both weak and strong ties. The current strategy was used to empirically tested networks among students in U.S. universities constructed by using online social networks. In most cases, the current strategy were found to be more effective in identifying influential spreaders than other methods, such as degree, betweenness, and k-shell decomposition strategy for the range of the set, and results were more accurate. Its effectiveness can be attributed to the neighbors identified by our strategy because the pattern of connections among neighbors was more striking than those produced using other strategies. For this reason, it can spread through the entire network more quickly and effectively. The current method kept its stability well throughout the missing processes, and has also been strongly performed robustness.

The performance of any strategy is influenced by the structure of networks. In the current method, when α exceeded the threshold level, such as when it approached 1, this was taken to indicate the highest probability of identifying the hub nodes in a single area. When it approached 0, that was taken to indicate that the method only preferred to choose bridge-hubs over communities. Both these conditions were unacceptable. Hence, finding the optimal values of the variable is still an unresolved issue.

**FIGURE LEGENDS**

**FIGURE 1**. **Visualizing the weak ties and strong ties in the network**. Nodes from two communities as illustrated by different logos, also illustrated are strong ties (solid) that source and target of the connection lie inside same community and weak ties (dash) between communities.

**FIGURE 2. Schematic illustration of strategies between k-shell decomposition and k-shell with**

**community**. The schematic representation of a network under the k-shell decomposition strategy; all nodes in innermost core have same k-core index, even those which have weak ties (dashed line). However, in the process of the k-shell with community strategy, these nodes have same index for $k_{core}^{S}$ and different index for $k_{core}^{W}$, 2 for A, 3 for B, 1 for C, and 0 for D. Finally, the node B lies at the most important location of the entire network, that is, has highest k-score = 3.5 compare with others (3 for A, 2.5 for C and 2 for D), according to equation (1). Here, $\alpha = 0.5$.

**FIGURE 3. Cumulative distribution of the number of weak ties and strong ties in empirical networks**. For every network, community structure was detected using the method proposed by Vincent et al. (Vincent et al. 2008). The weak ties and strong ties were then identified, and the number of weak ties $k_w$ and strong ties $k_s$ emanating from each node were recorded to give the cumulative distribution. Empirical results are shown for Caltech (squares), Princeton (circles), Georgetown (upward-facing triangles), and the University of Oklahoma (downward-facing triangles).

**FIGURE 4. Roles and regions in the Z-P space for empirical networks**. Here, we classified node with $Z \geq 2.5$ as local hub and node with $Z < 2.5$ as normal node according to a previous work (Guimera and Amaral 2005), nodes can be naturally assigned into regions: (1) $P < 0.2$; (2) $0.2 \leq P < 0.6$; (3) $0.6 \leq P < 0.8$; (4) $P \geq 0.8$.

**FIGURE 5. Comparison of spreading efficiency of identification in empirical networks.** The difference in imprecision $\Delta \varepsilon_i$, $i$ denoting degree (left panel), betweenness (middle panel), and k-core (right panel), are shown for each networks as function of $f$. The positive percentages shows that k-score is more accurate. Results are obtained by averaging over 2000 for each node. Here, $\alpha_{Caltech} = 0.95$, $\alpha_{Princeton} = 0.95$, $\alpha_{Georgetown} = 0.88$, and $\alpha_{Oklahoma} = 0.96$ for networks, respectively.

**FIGURE 6. Comparison of average number of weak ties and strong ties among neighbors of nodes**

**identified by k-score with those others.** The difference $\Delta\langle k_w\rangle(\Gamma_i)$ in weak ties (similar to $\Delta\langle k_s\rangle(\Gamma_i)$ in strong ties) with $i$ denoting degree (circles), betweenness (upward-facing triangles) and k-core (squares) of neighbor set $\Gamma$ are shown for different $f$ in empirical networks. Here, $\alpha_{Caltech} = 0.95$, $\alpha_{Princeton} = 0.95$, $\alpha_{Georgetown} = 0.88$, and $\alpha_{Oklahoma} = 0.96$ for networks.

**FIGURE 7. Robustness of the strategy in networks with noise as modeled by random removal of connections.** $\tau$ is shown as the function $f$ of number of connections randomly removed from networks. Results are obtained by averaging over 2000 realizations for each value of removed connections. Here, $\alpha_{Caltech} = 0.95$, $\alpha_{Princeton} = 0.95$, $\alpha_{Georgetown} = 0.88$, and $\alpha_{Oklahoma} = 0.96$ for networks.

# TABLES

| Networks | N | E | <k> | $k_{max}$ | <d> | C | r | H | G | Q |
|---|---|---|---|---|---|---|---|---|---|---|
| Caltech | 571 | 7127 | 24.963 | 96 | 2.965 | 0.574 | 0.001 | 0.363 | 9 | 0.794 |
| Princeton | 3975 | 23457 | 11.802 | 129 | 4.721 | 0.321 | 0.412 | 0.433 | 30 | 0.740 |
| Georgetown | 6309 | 73022 | 23.148 | 311 | 4.212 | 0.258 | 0.242 | 0.457 | 16 | 0.683 |
| U. Oklahoma | 6850 | 152985 | 44.667 | 247 | 4.361 | 0.524 | 0.488 | 0.511 | 42 | 0.926 |

**TABLE 1. Structural properties of networks**. Structural properties including the network size (*N*), number of edges (*E*), average degree (<*k*>), max degree ($k_{max}$), average shortest path length (<*d*>), clustering coefficient (*C*), degree assortativity (*r*), degree heterogeneity (*H*), the number of communities (*G*), and modularity (*Q*) are tabulated for each networks. These networks include data from the California Institute of Technology, Princeton University, Georgetown University, and the University of Oklahoma.

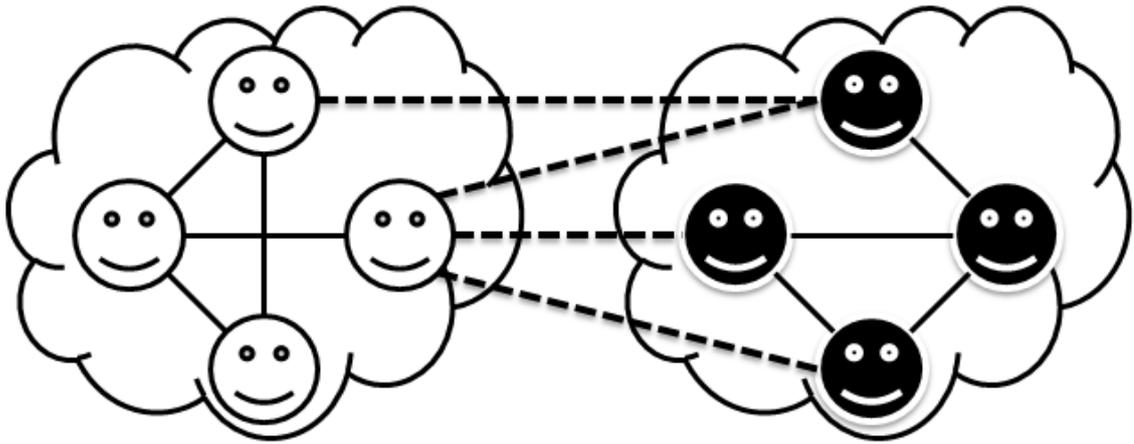

**FIGURE 1**. **Visualizing the weak ties and strong ties in the network**.

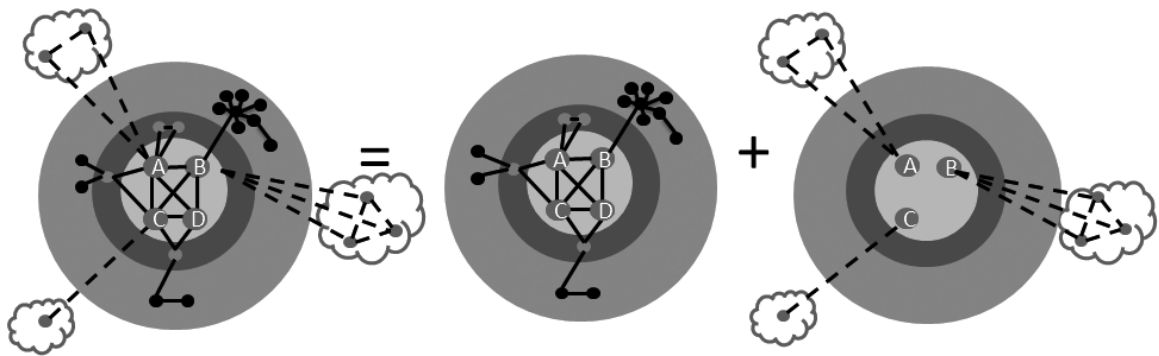

**FIGURE 2.** Schematic illustration of strategies between k-shell decomposition and k-shell with community.

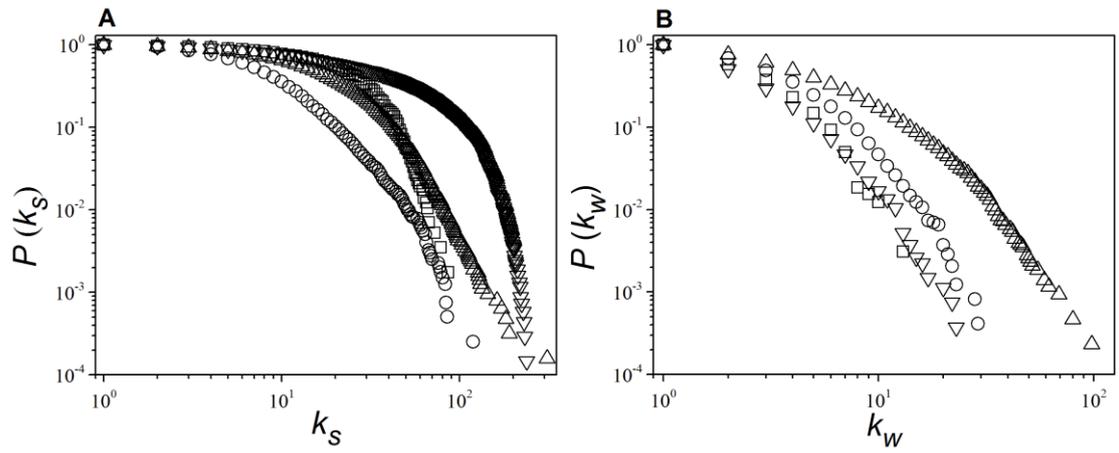

**FIGURE 3**. **Cumulative distribution of the number of weak ties and strong ties in empirical networks.**

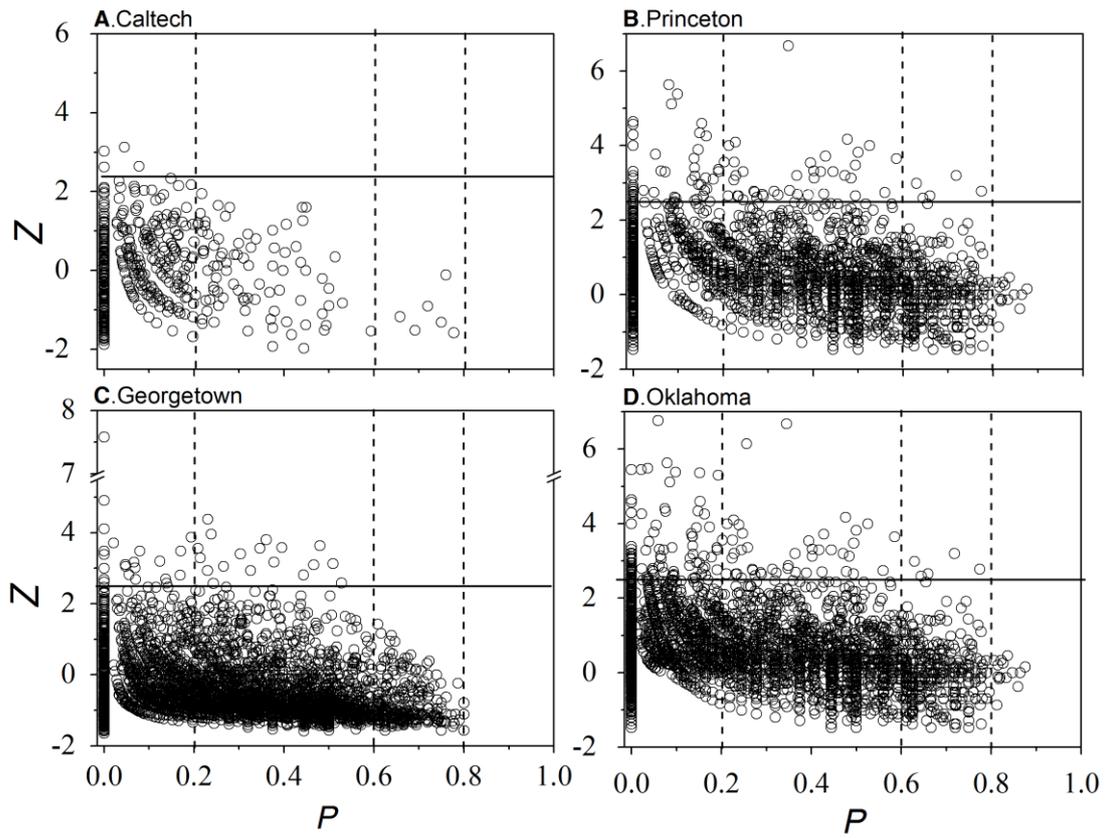

**FIGURE 4. Roles and regions in the *Z-P* space for empirical networks.**

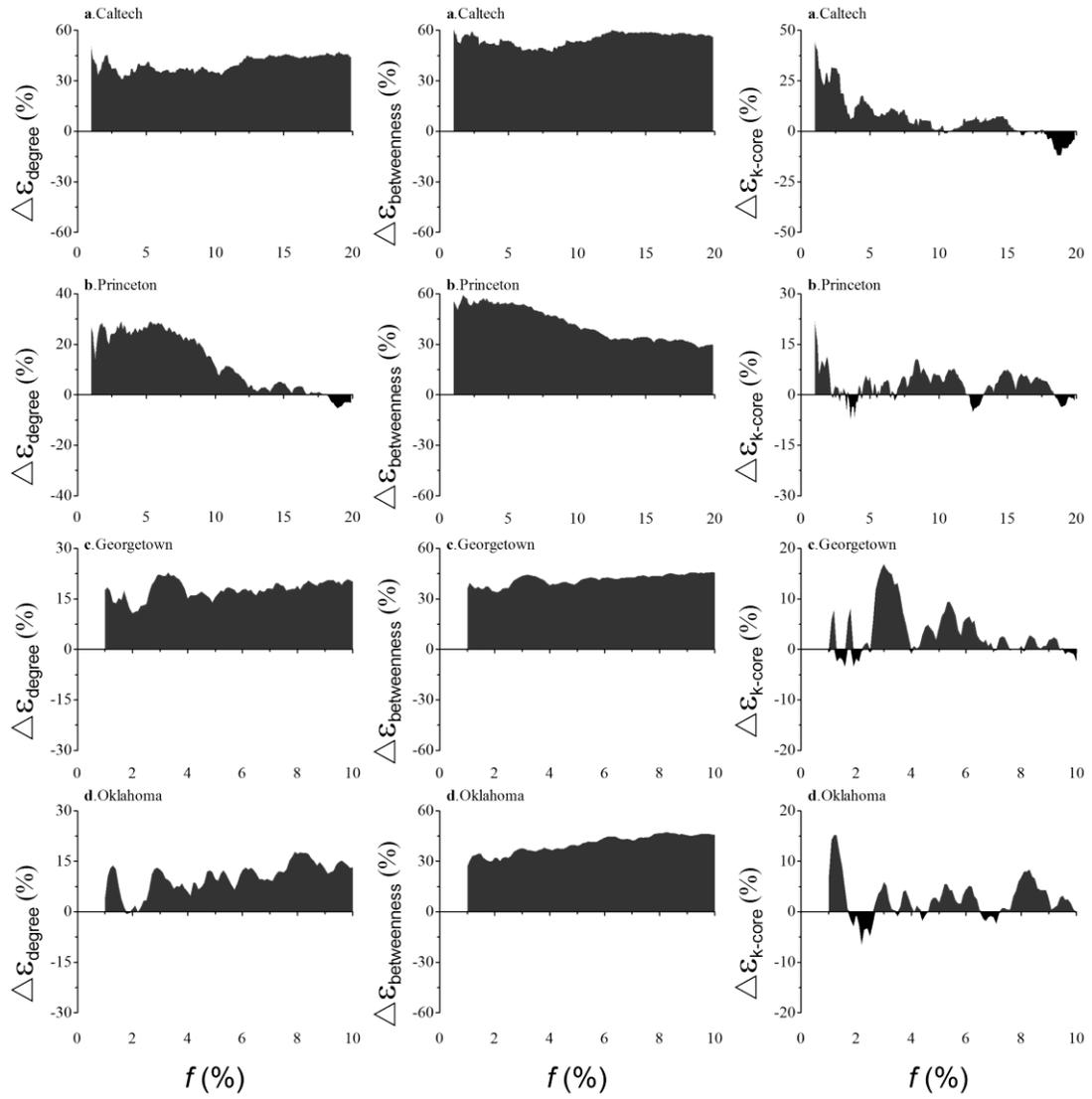

**FIGURE 5. Comparison of spreading efficiency of identification in empirical networks.**

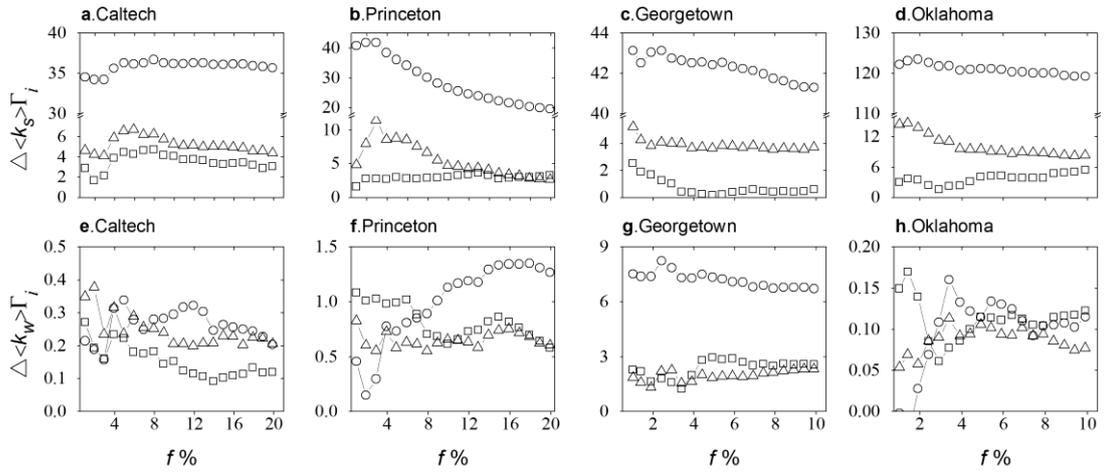

**FIGURE 6.** Comparison of average number of weak ties and strong ties among neighbors of nodes identified by k-score with those others.

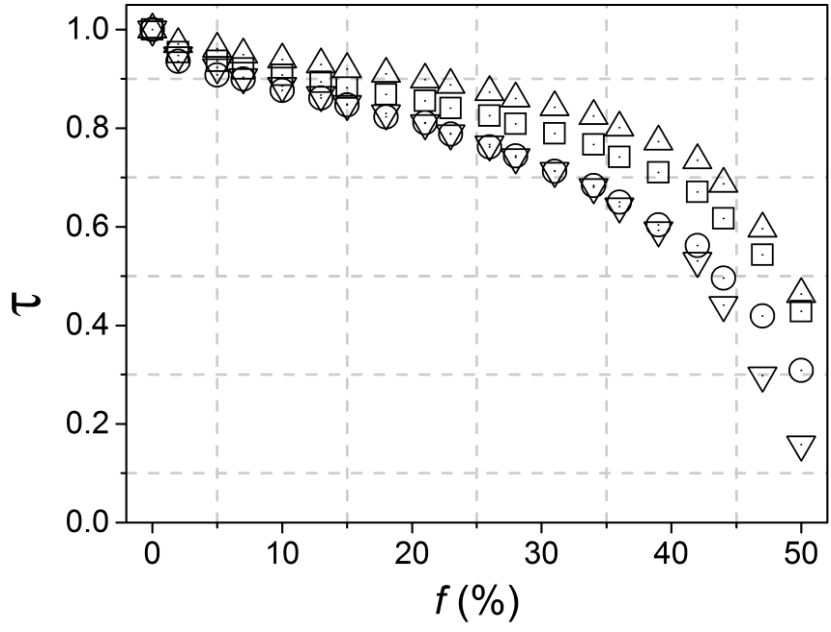

**FIGURE 7.** Robustness of the strategy in networks with noise as modeled by random removal of connections.